\documentclass[]{aa}

\usepackage{graphicx}

\begin{document}

\title{On the thermal behaviour of small iron grains}
\author{J\"org Fischera}

\offprints{J\"org Fischera}
\institute{Research School of Astronomy \& Astrophysics, Institute of Advanced Studies, The
Australian National University, Cotter Road, Weston Creek, ACT 2611, Australia}

\mail{fischera@mso.anu.edu.au}
\date{Received  ---/ Accepted ---}

 \abstract{
The optical properties of small spherical iron grains are derived using a Kramers-Kronig-consistent 
model of the dielectric function including its dependence on temperature 
and size. Especially discussed is the effect of the size dependence, which results
from the limitation of the free path of the free electrons in the metal 
by the size of the grain, on the
absorption behaviour of small iron spheres and spheroids. The estimated absorption
properties are applied to study the temperature behaviour of spherical and spheroidal grains 
which are heated by the interstellar radiation field.
}

\maketitle


\section{Introduction}

{ According to} the cosmic abundances { given by Anders \& Grevesse} (\cite{Anders89}),
iron is the ninth most abundant element in the interstellar medium (ISM) 
and should therefore be an essential
constituent of interstellar dust grains. { Iron} particles are expected to form independently of the oxygen-to-carbon ratio in circumstellar shells (see Jones \cite{Jones90}; and references therein)
and have been 
supposed to be, apart from silicates, the main condensates in the metal-rich core region
of supernova ejecta (Hoyle \& Wickramasinghe \cite{Hoyle70}). \object{SN 1987A} was the first
supernova where the condensation of new grains in ejected material
could be observed. The composition of these grains could not be clarified but indications, 
summarised by Wooden (\cite{Wooden97}), point to predominantly iron-rich grains.

Even though the ISM will probably be enriched in iron particles it is uncertain {whether}
they can survive and play an important role as a dust component of
interstellar grains (Duley \cite{Duley80}, Jones \cite{Jones90}). 
If they are present, iron particles may affect the interstellar extinction 
(Wickramasinghe \& Nandy \cite{Wickramasinghe71}), the
chemistry in the ISM due to catalytic processes on grain surfaces
(Tabak \cite{Tabak78}), and may contribute to the diffuse 
emission from interstellar dust grains (Chlewicki \& Laureijs \cite{Chlewicki88}). 
Further, iron grains may be an important grain species 
to explain the Rosseland mean opacity in molecular clouds cores (above $\approx 575$~K)  and accretion disks (above $\approx 425$~K) 
(Pollack et al. \cite{Pollack94}).
{On the other hand it has been claimed, based on observations at 90 GHz,
that not more than 5\% of interstellar iron is in form of
metallic iron grains or inclusions (Draine \& Lazarian, \cite{Draine99}).
This would support the hypothesis that the survival time of a relatively pure iron grain is 
quite short. However, even though overall iron may be not an important constituent 
of interstellar grains it does not exclude the possibility of local enhancements where
these grains may form.}
 

Mathis, {Rumpl} \& Nordsieck (\cite{MRN77}) found, from analysis of the extinction curve, that the interstellar medium is mainly composed of grains much smaller than $\sim 0.25~\mu{\rm m}$. Very small grains are furthermore thought to be responsible for the diffuse IR emission at wavelengths shorter than $\sim 60~\mu{\rm m}$ (see e.g. Draine \& Anderson \cite{Draine85}) due to the stochastic heating process, where very small grains reach temperatures much higher than the corresponding equilibrium temperature. To model their emission or their ir optical properties, it is usually assumed that the dielectric function derived for solids can also be applied for very small grains. At least for small metallic grains, as shown by experiments with very small silver spheres, it is known that the dielectric function has to be corrected for grain size (Kreibig \& Fragstein \cite{Kreibig69}; Kreibig \cite{Kreibig74}). The influence of the grain size on the dielectric function can be even stronger than its dependence on temperature. One expects that the optical properties of small metallic spheres should be much less affected by temperature than those of larger grains as shown, for example, by calculations of the absorption behaviour of spherical graphite grains (Draine \& Lee \cite{DraineLee84}). 

{Here, we} present calculations for the temperature behaviour of pure iron grains in the ISM heated by the interstellar radiation field (ISRF) where we have taken into account both the temperature and the size dependence of the dielectric function. It has been suggested that a part of the interstellar grains may have a non-spherical shape (see Voshchinnikov et al. \cite{Vosh99}). Therefore, apart from simple iron spheres, spheroidal iron grains will also be considered. The influence of the shape of small grains of different compositions, including iron, on temperature has also been discussed by Voshchinnokov et al. (\cite{Vosh99}). Their results are based on the dielectric function of solids using a sophisticated code to derive the optical properties of spheroidal grains. Here a much {simpler} approach is used demonstrating the effect of the size dependence of the dielectric function on the cooling behaviour of spheroidal iron grains.

In Sect.~\ref{model} we describe the model used to derive the optical properties of small iron spheres and analyse in more detail the absorption behaviour of small spherical and spheroidal iron grains. In Sect.~\ref{iron_ISM} the optical properties are used to derive the temperature behaviour of spherical and spheroidal small iron grains in the ISM which are heated by the ISRF. The derived temperature distributions are used to obtain the emission spectrum of spherical iron grains assuming a grain size distribution as proposed for the ISM including the stochastic heating process of very small grains. The results are discussed in Sect.~4 and summarised in Sect.~5.

\section{The optical properties of iron grains}
 
\label{model}
The interaction of an electromagnetic wave of a given energy with a grain
is generally determined by the response of the corresponding solid, expressed by
the complex dielectric function $\epsilon$, the shape, and the size of the grain. 

As for non-metallic grains the magnetic permeability is taken to be $\mu=1$. As will also be shown
this assumption should be sufficient to describe the cooling of iron grains in the ISM, 
which is the main purpose of this paper.
The contribution of the magnetizability of metallic grains to the optical properties
has been studied by Draine \& Lazarian (\cite{Draine99}) and is
only significant at wavelengths larger than $\sim 1000~\mu{\rm m}$.

\subsection{A model for the dielectric function}

  \label{model_dielectrfunct}
  To derive the optical properties of spherical iron grains we used
  a self consistent simple model for the complex dielectric function
  $\epsilon(\omega)=\epsilon_1(\omega)+i\epsilon_2(\omega)$ based on optical data
  derived for bulk iron at a temperature of $298~\mathrm{K}$.

{We followed basically the method used by Draine \& Lee (\cite{DraineLee84}) to determine the optical properties of small graphite grains} and separated the dielectric function in $\delta \epsilon^f$ and $\delta\epsilon^b$, which correspond to the free and the {bound} electrons of the grain:
  \begin{equation}
    \epsilon=1+\delta\epsilon^f+\delta\epsilon^b.
  \end{equation}
It is assumed that the dependence on temperature and grain size is given through the free electron part which is described by a single Drude term
  \begin{equation}
    \label{drudemodell}
    \delta\epsilon^f=\frac{-\omega_p^2}{\omega^2+i\Gamma\omega}
  \end{equation}
where $\Gamma$ is the collision rate and $\omega_p$ the plasma frequency. 

The parameters of the Drude term of bulk iron are determined at long wavelengths where the free electron gas dominates the optical behaviour of metals. To derive a self consistent dielectric function we have chosen the imaginary part as given by $\epsilon_2^b=\epsilon_2-\delta\epsilon_2^f$ and used the Kramers-Kronig relation (Landau \& Lifschitz, \cite{Landau85}) to estimate the real part $\epsilon_1^b$.
 
{The main dependence of $\delta \epsilon^f$ on size and temperature is given by the collision rate $\Gamma$ which can be taken as the sum of the collision rates $\Gamma _{\rm i}$ and $\Gamma_{\rm s}$ due to scattering events in the solid and at the outer surface of the grain (Kreibig \& Fragstein \cite{Kreibig69}; Kreibig \cite{Kreibig74}).  A further temperature dependence is assumed to  be caused by the thermal expansion of the solid which alters the electron density $n_{\rm e}$ and therefore $\omega_{\rm p}$ ($\omega_{\rm p}\propto n_{\rm e}^{1/2}$).
  
Inside the solid, electrons are scattered at impurities, defects and phonons, but only the phonon scattering events lead to a strong temperature dependence. In the case of bulk iron, the scattering at impurities and defects is negligible with the exception of very low temperatures where they are the dominant scattering process inside the grain.
  
As a first approximation the scattering rate at phonons may be assumed to be independent of the amount of impurities and defects and the scattering rate at impurities and defects independent on temperature so that the effects can be easily de-coupled. Under these assumptions the collision rate due to impurities and defects can be scaled by a factor $\zeta$ to allow consideration of the cases where the iron grains are either less pure or have more defects in comparison to pure iron grains.
  
 \begin{figure}[htbp]
    \includegraphics[width=\hsize]{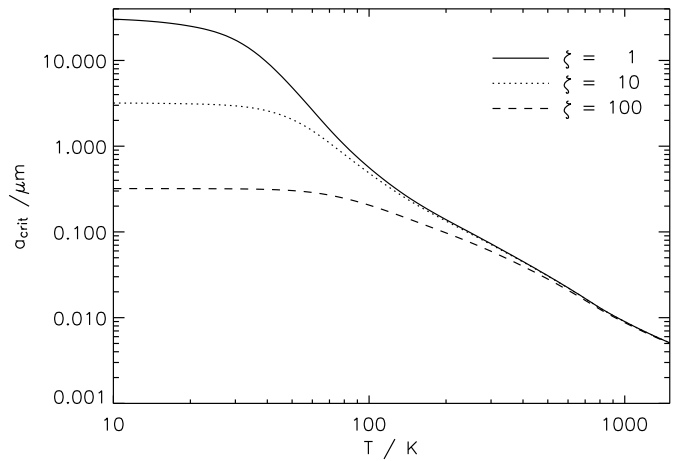}
  	\caption{\label{crit_size}
	Critical grain sizes of iron spheres where the collision rate $\Gamma_{\rm i}$ of the free
 	electrons in the material is equal to the collision rate at the surface shown for $\zeta=1$,
	 $10$, and $100$. Below the curves the collision rate is dominated by surface scattering.}
  \end{figure}
 
The total collision rate is given by
  \begin{equation}
    \label{Gamma}
     \Gamma = \Gamma_{\rm i}+\Gamma_{\rm s}=
     \frac{\omega_p^2(T)}{4\pi\sigma_0(T,\zeta)}+\frac{v_F}{\beta a}
  \end{equation}
where $\sigma_0(T,\zeta)$ is the conductivity of the iron grains taken to be
     \begin{equation} 
     \frac{1}{\sigma_{0}(T,\zeta)}=\frac{1}{\alpha}\left\{\frac{1}{\sigma_0^{\rm bulk}(T)}+\frac{\zeta-1}{\sigma_0^{\rm bulk}(0~{\rm K})}\right\}.
     \end{equation}
$\sigma_0^{\rm bulk}(T)$ is the adopted temperature-dependent conductivity of bulk iron and $\alpha = \sigma_0(298~{\rm K})/\sigma^{\rm bulk}_0(298~{\rm K})$ a (small) correction factor where $\sigma_{0}(298~{\rm K})$ is the conductivity obtained by fitting the Drude term to the optical properties of bulk iron (see below). $v_F$ is the Fermi velocity of the electrons and $a$ the radius of the grain. The parameter $\beta$ is of order one and is given by the scattering process at the surface. As Draine \& Lee we have taken $\beta=1$, which is equivalent to isotropic scattering. 
     
  The collision rate at the
  surface becomes important when the free path of the electrons is greater than the size of 
  the grain. Where the collision rate in the solid is equal to the collision rate at 
  the surface is shown in Fig.~\ref{crit_size} for different values $\zeta$.
  In the case of pure very cold iron grains ($\zeta=1$)
  the scattering should be dominated by surface scattering for all grains with
  radii smaller than $\sim 20~\mu{\rm m}$. If the grains are less pure and have more defects
  the effect is limited to smaller grain sizes. 
  
As the purity of `astrophysical' iron grains is not known we considered pure iron grains. The collision rate assumed in this paper can therefore be regarded as a lower limit of the collision rate of actual iron grains. However, this assumption should not affect interstellar iron grains which, if they do exist, are believed to be quite small with maximum sizes not much larger than $\sim 0.1~\mu{\rm m}$. As seen in Fig.~\ref{crit_size}, their mean free path should be determined by the size of the grain unless the grains are highly disordered or impure with $\zeta>100$. The effect of the purity and defects on the temperature behaviour of larger iron grains will be discussed in Sect.~\ref{sectdiscuss1}.}

    \begin{figure*}[htbp]
    \includegraphics[width=0.49\hsize]{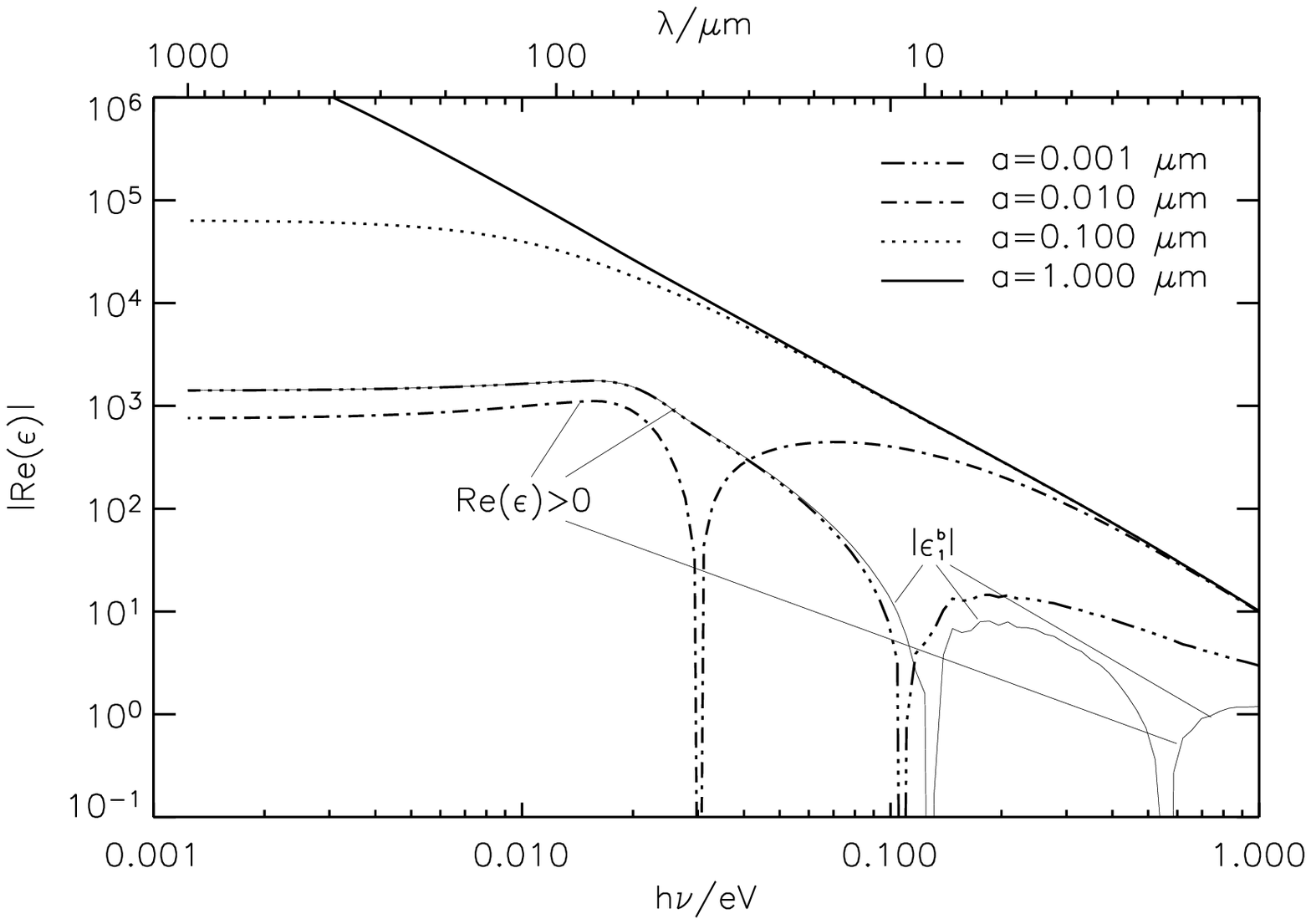}
    \hfill
    \includegraphics[width=0.49\hsize]{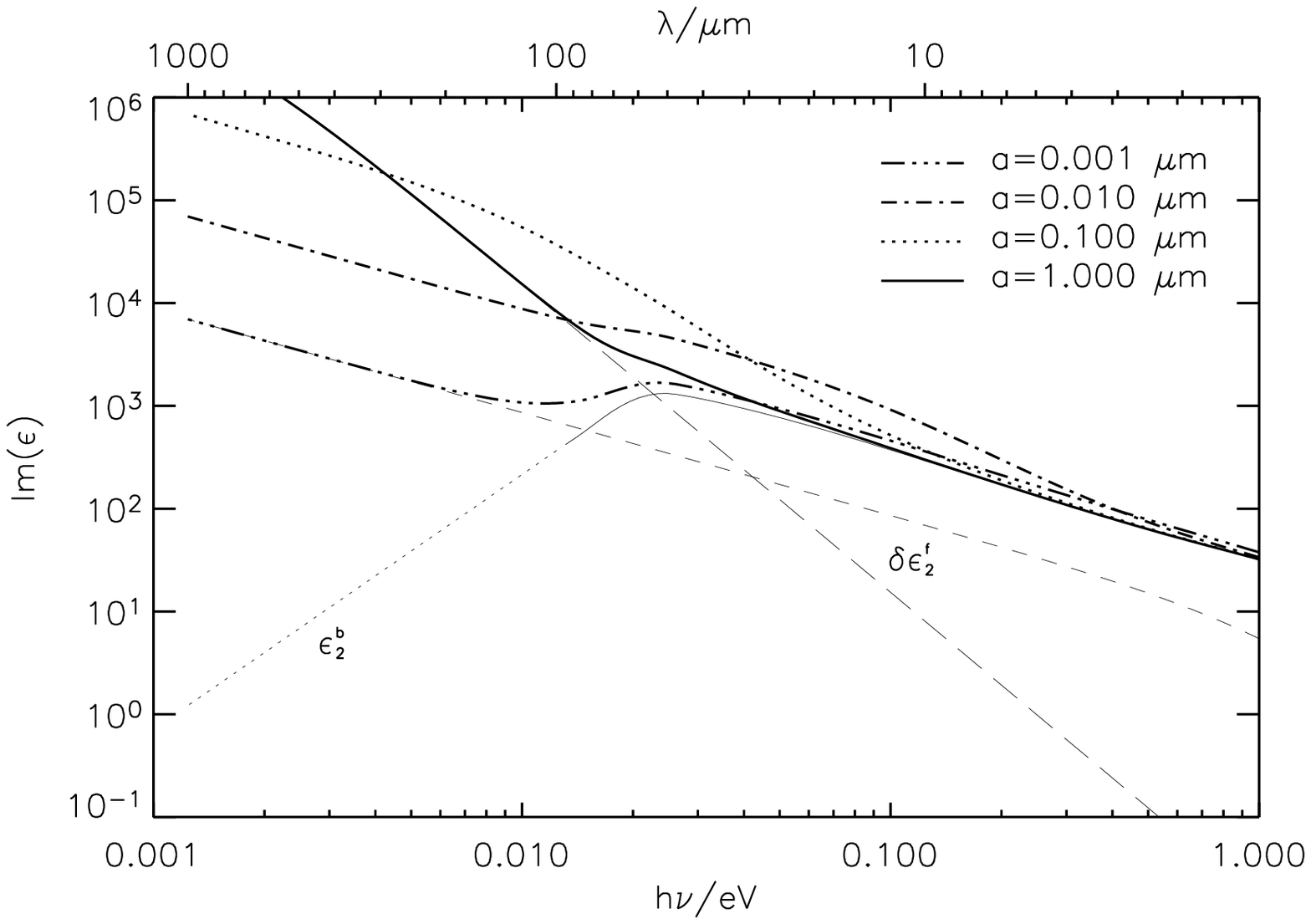}
    \caption{
      \label{epsilon_ir}
  {Real and imaginary part of the dielectric function for individual pure small iron spheres at energies 
  above 1~eV. For comparison also the real and imaginary parts of $\epsilon^{\rm b}$ are shown as thin solid lines. The part of $\epsilon_2^{\rm b}$ shown as thin dotted line are extrapolated values used for the
  Kramers-Kronig relation to derive $\epsilon_1^{\rm b}$. The real part of the dielectric
  function is always negative except for the three branches shown. The short and the long dashed thin
  lines in the right hand figure refer to the dielectric function 
  $\delta\epsilon_2^f=\mbox{Im}(\epsilon-\epsilon^{\rm b})$ of an iron sphere with 
  $a=0.001~\mu{\rm m}$ and $a=1.0~\mu{\rm m}$, respectively.}}
  \end{figure*}
  \begin{figure*}
    \includegraphics[width=0.492\hsize]{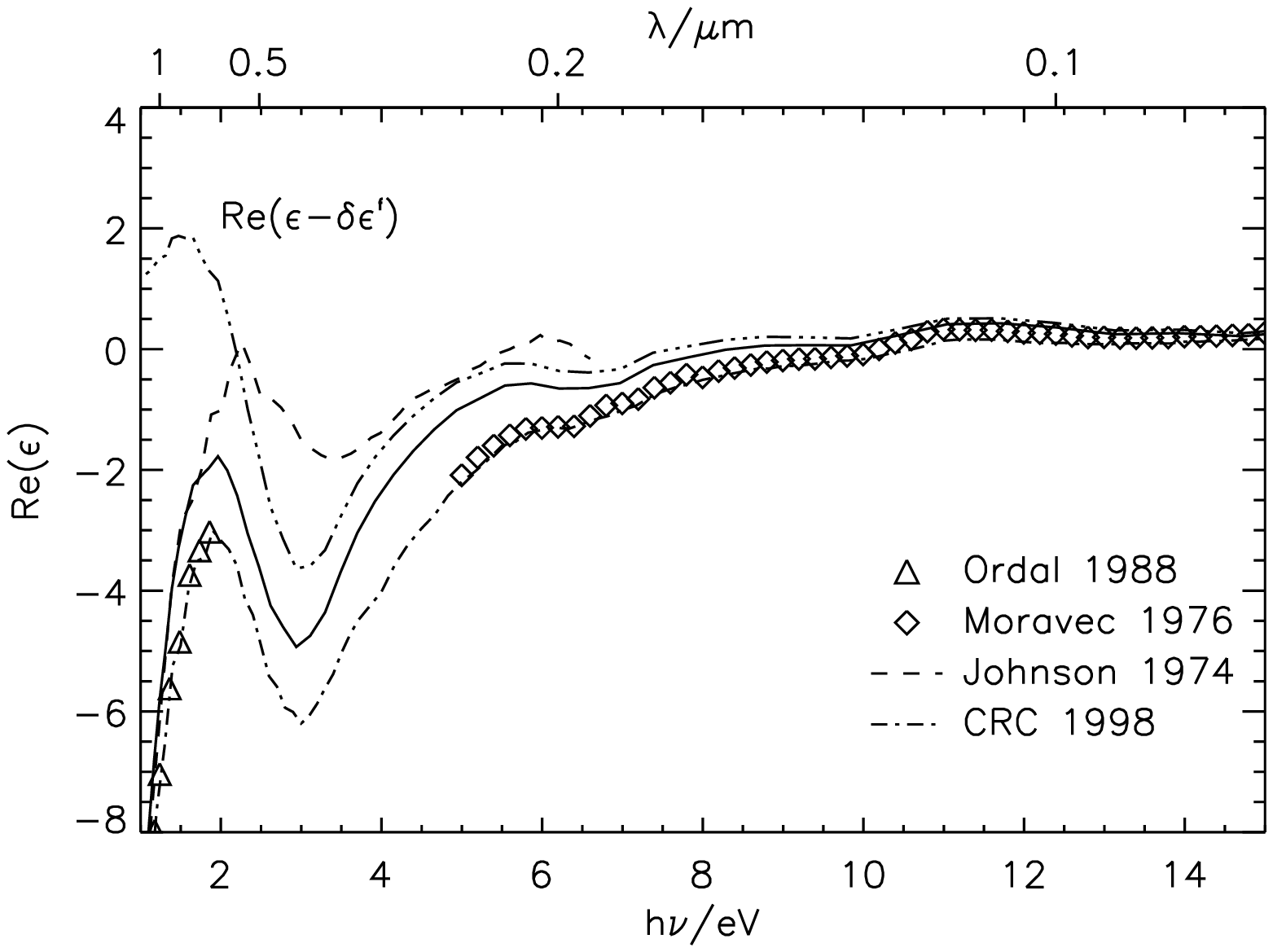}
    \hfill
    \includegraphics[width=0.48\hsize]{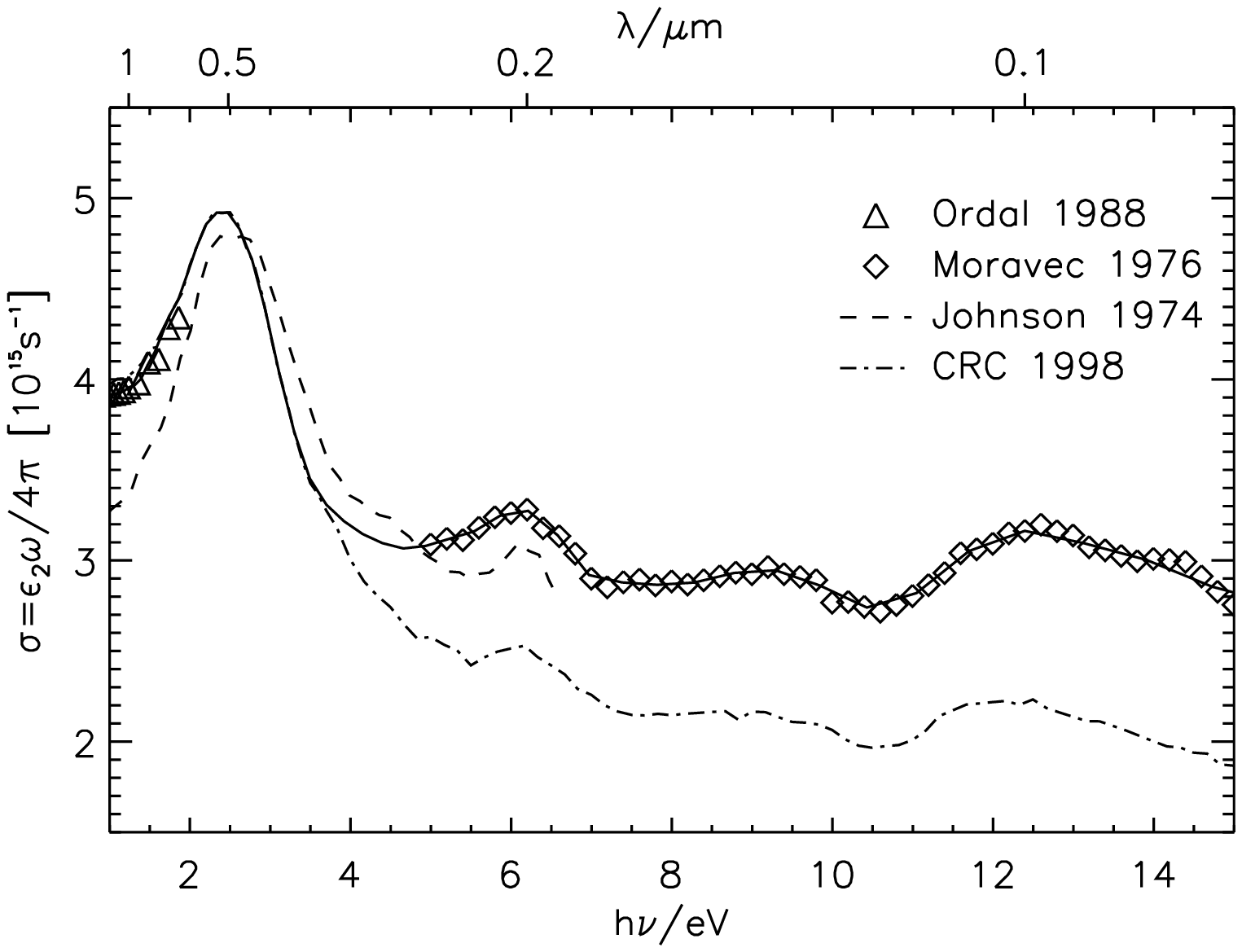}
    \caption{
      \label{epsilon_optuv}
      Real and imaginary part, given as conductivity
    $\sigma=\epsilon_2\omega/4\pi$, of the dielectric function $\epsilon$ above
    1eV at a temperature of 298~K. The chosen values of the imaginary part and the 
    result for the real part after Kramers-Kronig analysis are shown as solid lines. 
    The values differ somewhat from the published values. The derived values of
    the real part of the dielectric function lie between those from
				Johnson \& Cristy (\cite{Johnson74}) and those from Moravec et al. (\cite{Moravec76}).
    The real part $\epsilon_1^b={\rm Re}(\epsilon-\delta\epsilon^f)$ (shown as dashed-dotted line)
    is the direct result derived with the Kramers-Kronig relation from the imaginary 
				part $\epsilon^b_2=\epsilon_2-\delta\epsilon_2^f$.}
  \end{figure*}

The optical data used to derive the optical properties can be taken from Table \ref{taboptconst}. We interpolated data from Lynch \& Hunter (\cite{Lynch96}) and Moravec et al. (\cite{Moravec76}) in the energy range 51.57~eV to 26~eV using $f(E)=cE^\gamma$.  A smooth transition was made {in the range $3.5$ to $5$~eV} between data from Moravec et al. and the data published in the `CRC Handbook for Chemistry and Physics' (\cite{CRC98}).
  Above 0.8~eV we used the averaged data from Ordal et al. (\cite{Ordal88}).

  \begin{table}[ptbh]
    \caption{Used optical data}
    \label{taboptconst}
          \begin{tabular}{c|l}
       \hline
       \hline
        energy intervall [eV] & reference \\
          \hline
          $10\,000-51.57$ & Lynch \& Hunter (\cite{Lynch96}) \\
          $51.57-26.0$ & interpolation (see text)        \\
          $26.-5$ & Moravec et al. (\cite{Moravec76}) \\
          $3.5-1.55$ & CRC  (\cite{CRC98}) \\
          $1.55-0.0124$ & Ordal et al. (\cite{Ordal88}) \\
          $0.0124-0.00124$ & extrapolation (see text)
          \end{tabular}
  \end{table}

From the optical data in the range of $1$ to $100~\mu{\rm m}$ we determined the plasma frequency $\omega_p$ and the internal collision rate $\Gamma_{\rm i}$ of bulk iron at room temperature which are given in Table~\ref{drudeparameter}. To prevent amplification of the electromagnetic wave inside the solid we have taken care that $\epsilon_2^b=\epsilon_2-\epsilon_2^f>0$ in the considered energy range. The derived parameters of the Drude term ($\omega_p=27\,020~{\rm cm^{-1}}$, $\Gamma_{\rm i}=125.2~{\rm cm^{-1}}$) are slightly smaller than the ones used by Ordal et al. for extrapolation above $100~\mu{\rm m}$ ($\omega_p=29\,500~{\rm cm^{-1}}$ and $\Gamma_{\rm i}=156~{\rm cm^{-1}}$) but very close to those used by Draine \& Lazarian (\cite{Draine99}).

  The DC conductivity of iron at room temperature is found to be 
  $\sigma_0=\omega_p^2/4\pi\Gamma_{\rm i}=7.65\cdot 10^{16}\,{\rm s^{-1}}$
  which is only about a factor of $1.2$ lower than the measured value of $9.12\cdot 10^{16}~{\rm s^{-1}}$ 
  (CRC \cite{CRC98}). For comparison
  in the case of graphite the DC conductivity given by the parameters of the Drude term is a factor of 3 smaller 
  than the measured value (Draine \& Lee \cite{DraineLee84}). 

  \begin{table}[hbtp]
    \begin{minipage}[]{\hsize}
    \caption{Parameters used for the dielectric function $\delta \epsilon^{\rm f}$
    \label{drudeparameter}}
    \renewcommand{\thefootnote}{\it\alph{footnote}}
    \begin{tabular}{ll|c}
       \hline
       \hline
        \multicolumn{2}{c|}{parameter} & value \\
        \hline
        plasma frequency (298K)\footnote[1]{Fit parameters with single
    Drude term.}& $\omega_p$ [s$^{-1}$] & $5.090\cdot 10^{15}$ \\
        internal collision rate (298K)\footnotemark[1]&
        $\Gamma_\mathrm{i}$ [s$^{-1}$] & $2.693\cdot 10^{13}$ \\
        Fermi velocity\footnote[2]{Taken from Ashcroft \& Mermin (\cite{Ashcroft76}).} & 
								$v_F$  [cm/s] & $1.98\cdot 10^8$
    \end{tabular}
    \end{minipage}
  \end{table}

  The derived dielectric function is shown in Fig. \ref{epsilon_ir} and Fig. \ref{epsilon_optuv}.
  At long wavelengths (Fig. \ref{epsilon_ir}) the functional form of $\epsilon^b$ 
  is qualitatively similar to the dielectric function $\epsilon_{\perp}^b$ of graphite perpendicular
  to the main axis (Draine \& Lee \cite{DraineLee84}), which shows metal-like behaviour.
  The imaginary part $\epsilon_2^b$ increases 
  with wavelength and reaches its highest values at $\sim 60~\mu{\rm m}$. Beyond $100~\mu{\rm m}$ we
  assumed that the values of $\epsilon_2^b$ decrease with a power law.
  The explanation of the behaviour of the dielectric function $\epsilon^b$ in the IR 
		is out of the scope of this paper and it cannot be excluded that this behaviour is 
  a result of the simplification in modelling the free electron regime with a single Drude term.

{The temperature dependent DC conductivities $\sigma^{\rm bulk}_0(T)$ of bulk iron are taken from the `CRC Handbook of Chemistry and Physics' (\cite{CRC98}).} The expansion coefficients of iron for correcting the plasma frequency with temperature was taken from the `American Institute of Physics Handbook' (\cite{AIP72}). For the collision rate at the surface we used for the free electron gas of iron a Fermi velocity $v_F=1.98\cdot 10^8~{\rm cm/s}$ (Ashcroft \& Mermin \cite{Ashcroft76}).

\subsection{Optical properties of spherical pure iron grains}

  \label{optprop_spheres}
\begin{figure}
    \includegraphics[width=\hsize]{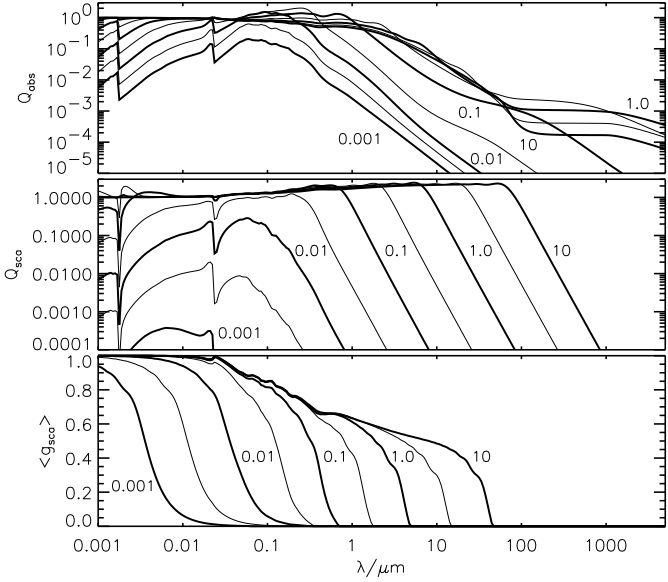}
    \caption{\label{optprop}
      Derived optical properties of spherical iron grains. Shown are the absorption
      coefficients $Q_\mathrm{abs}(\lambda,a)$, scattering coefficients $Q_\mathrm{sca}(\lambda,a)$,
      and the scattering parameter $g_\mathrm{sca}(\lambda,a)=\left<\cos(\theta)\right>$ at 25~K for various grain radii 
      from $10~\AA{}$ to $10~\mu{\rm m}$ with a difference in radius of $\Delta\log{a[\mu{\rm m}]}=0.5$.
      Thick lines are labeled with the corresponding grain radii in microns.}
  \end{figure}

We estimated the optical properties of spheres above $2\pi a/\lambda =0.1$ 
using the MIE-formalism from Bohren \& Huffman 
(\cite{Bohren83}). At longer wavelengths we used the dipole approximation for spheres presented by Draine \& Lee (\cite{DraineLee84}). The results for iron grains at a temperature of 25~K are shown in Fig.~\ref{optprop}. At this temperature the mean free path of the electrons would be larger than the grains so that the collision rate $\Gamma$ is given by the collision rate at the grain surface and therefore through $\Gamma \approx v_F/a$.  While scattering ($Q_{\rm sca}$, $g_{\rm sca}$) is almost insensitive to the grain temperature, the absorption of larger iron grains can have large temperature variations. 

\begin{figure}
    \includegraphics[width=\hsize]{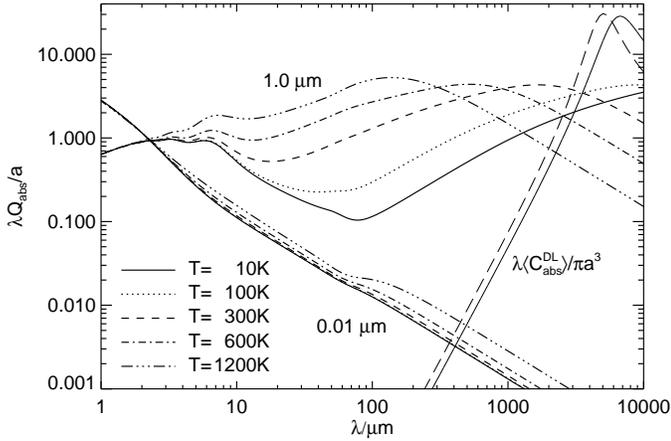}
    \caption{\label{abstemp}
      Effect of the temperature on the absorption behaviour of spherical iron grains.
      For various temperatures, the absorption coefficients of iron grains with radii
      1.0 and $0.01~\mu{\rm m}$ for $\mu=1$ (thick lines) are shown. For comparison the
      absorption coefficients due to magnetisation effects for magnetic single and 
      multi-domain iron grains well below the Curie temperature ($744^{\circ}~{\rm K}$) are
      given (thin dashed and solid line).}
  \end{figure}

The dependence of the absorption behaviour of iron grains on temperature is shown in Fig.~\ref{abstemp} for two different grain sizes with radii of 0.01 and $1~\mu{\rm m}$. {For comparison we also show the  absorption cross section $\left<C_{\rm abs}^{\rm DL}\right>$ due to magnetisation effects inside iron spheres well below the Curie temperature as derived by Draine \& Lazarian \cite{Draine99}. If correct this effect clearly dominates the absorption of cold iron grains at wavelengths longer than $\sim 1000~\mu{\rm m}$. Above the Curie temperature the iron spheres become non magnetic and the absorption at long wavelength should be determined by the electronic response of the material. As the magnetisation does not affect the cooling behaviour of the grains this effect will be ignored in the following.}
              
Due to the high Fermi velocity only the optical properties of larger iron grains with 
$a\gg 0.01\mu{\rm m}$ are affected by grain temperature. Even a variation of grain temperature over a huge range (10 - 1200~K) causes only a slight change in the absorption behaviour of grains with $a=0.01~\mu{\rm m}$. Grains with a radius of $a=1~\mu{\rm m}$ on the other hand have, at a wavelength of $100~\mu{\rm m}$, an order of 10 higher absorption  at 300~K than at 10~K. In this wavelength region the absorption coefficients increase with temperature.  At longer wavelengths it is possible that colder iron grains emit more readily than warm grains do. 

 \subsubsection{Dipole absorption of spherical pure iron grains}
   \label{abs_spheres}
{As is well known, metallic spheres show relatively complex absorption characteristics at long wavelengths. The behaviour is in contrast to non-metal grains where the IR-absorption is close to $Q_{\rm abs}\propto a/\lambda^2$. This is due to the fact that at long wavelengths the absorption behaviour of grains with a high dielectric function, as for metals, can be dominated by the magnetic dipole absorption caused by eddy currents which are created through the changing magnetic field inside the conductor (Landau \& Lifschitz \cite{Landau85}). In our calculation the absorption behaviour of iron spheres at wavelengths shorter than $\sim 60~\mu{\rm m}$ is additionally partly affected by the high values of the dielectric function $\epsilon^b$.
      
      However, as seen in Fig.~\ref{optprop} and Fig.~\ref{abstemp}, inside certain regions of the parameter space defined by the wavelength $\lambda$ and the grain size $a$ the absorption coefficient can still be described by a simple approximation ${\rm Q_{abs}}(\rm a,\lambda)\propto a^s\lambda^t$ with fixed values for $s$ and $t$. The absorption properties at long wavelengths can be explained using the electric and the magnetic dipole approximation for the cross sections $C_{\rm abs}=\pi a^2Q_{\rm abs}$ and are discussed in some detail in the Appendices \ref{approx} and \ref{idealised_dipole}.}
      

 \subsection{Dipole absorption of small spheroidal pure iron grains}

  \label{abs_spheroids}
As a special case of non-spherical iron particles we will consider spheroidal iron grains. In adopting the model of the dielectric function described in Sect.~\ref{model_dielectrfunct} to 
non-spherical grains one has to choose the appropriate value of the collision rate in Eq.~\ref{Gamma}.  As an approximation of the mean free path due to surface scattering we have taken the averaged distance from the center to the surface:
  \begin{equation}
   \left<r\right> =
    \frac{b}{e}\times \left\{
      \begin{array}{lll}
        \arcsin(e) & a>b &(\mathrm{prolate}), \\
        \ln\left(e+\sqrt{e^2+1}\right) & b>a &(\mathrm{oblate}),
        \end{array}
      \right.
  \end{equation}
where $e^2=|1-(b/a)^2|$. $a$ and $b$ are the semi-major axis parallel and vertical to the
rotation axis. For a sphere $\left<r\right>$ is equal to the mean free path of isotropic scattering 
($\beta=1$). In the limit of an elongated prolate grain ($e \approx 1$) the averaged distance 
from the centre to the surface is equal to $\pi b/2$ and only slightly different to the mean free path of $2b$ in the case of an infinite needle if isotropic scattering is assumed. 

In general for the calculation of the optical properties of non-spherical grains, even for spheroidal grains, can be very complicated (see e.g. Voshchinnikov \& Farafonov, \cite{Vosh93}). Here we are interested in the temperature behaviour of grains in the interstellar medium where the radiation is almost isotropic and unpolarised. In such cases it is sufficient to use only the cross section averaged over the orientations of the spheroids. 
  
  \begin{figure*}
    \includegraphics[width=0.49\hsize]{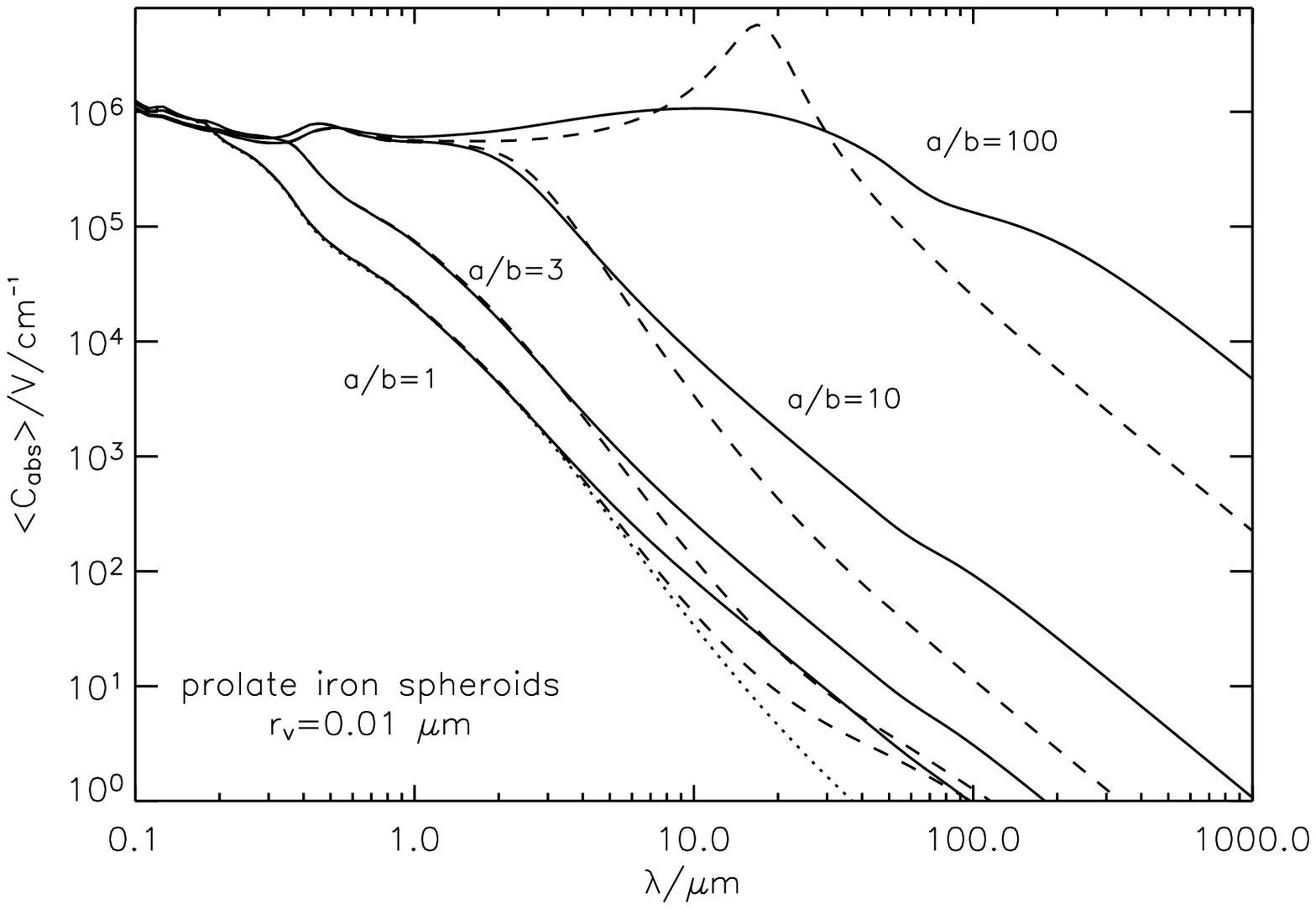}
    \hfill
    \includegraphics[width=0.49\hsize]{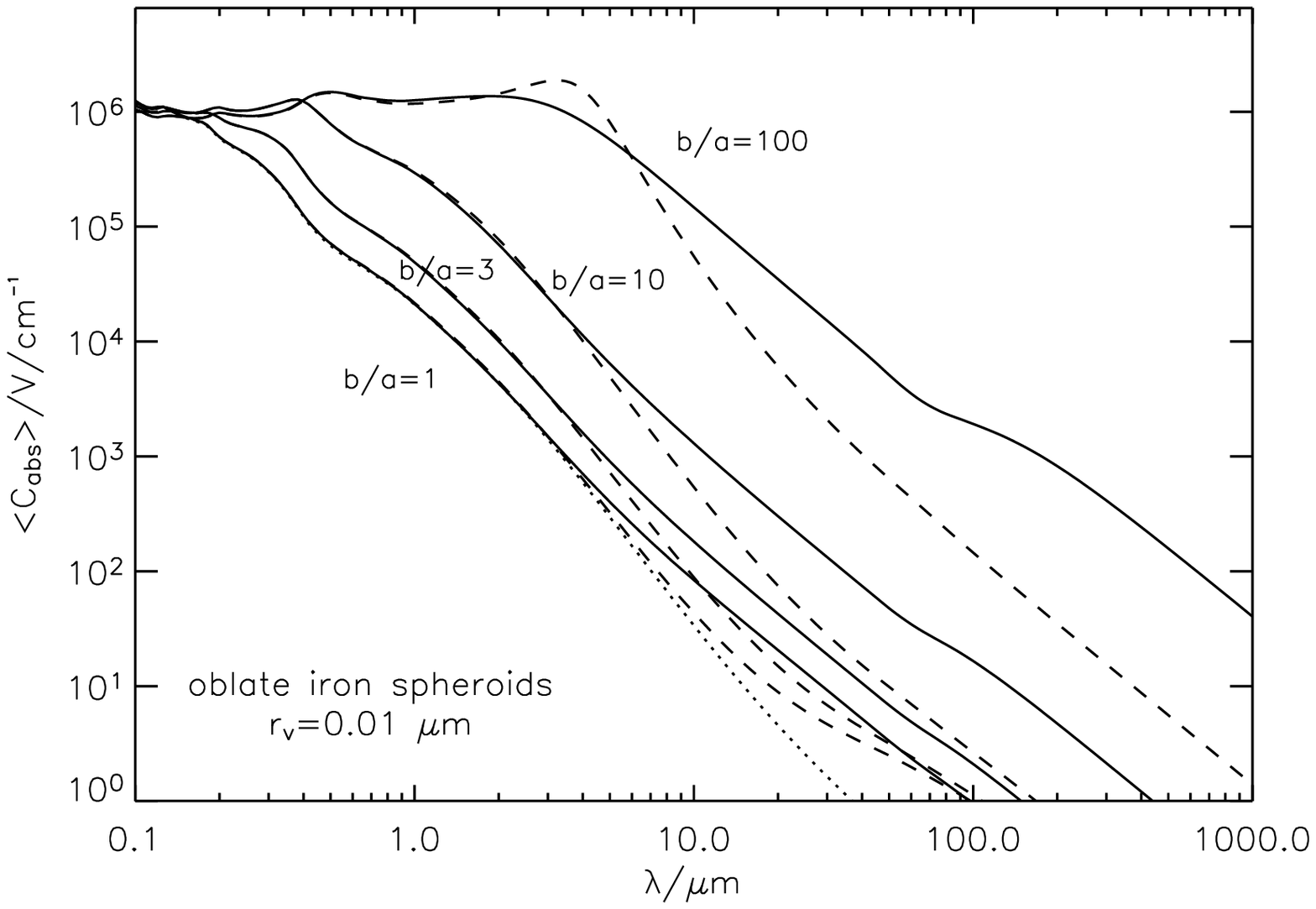}
    \caption{
      \label{fig_abs_spheroids}
      Angle-averaged absorption cross section $\left<C_{\rm abs}\right>$ of prolate ($a>b$) and oblate ($a<b$) 
      iron grains with volumes equivalent to the volume of a sphere with radius $r_v=0.01~\mu{\rm m}$.
      For comparison also the absorption of a sphere is given. 
      Solid and dashed lines correspond to absorption where the size and temperature dependence 
      of the dielectric function has been taken into account and has been ignored, respectively. 
      The dotted line shows the electric dipole absorption of a sphere
      for the case of temperature and size independent dielectric function.}
  \end{figure*}
  
For simplicity, we considered spheroidal grains small in comparison to the wavelength of the absorbed photon so that the optical properties can be described by the dipole approximation. For the electric dipole absorption we used the approximation for ellipsoidal grains and for the magnetic dipole absorption we adopted the expression given for spheres (Draine \& Lee, \cite{DraineLee84}). 
  
The angle averaged absorption cross sections of spheroidal iron grains for a number of different ratios $a/b$ are shown in Fig.~\ref{fig_abs_spheroids} (solid lines). The volume of the spheroids is chosen to be that of a sphere with a radius of $r_v=0.01~\mu{\rm m}$.  For comparison also the cross sections for temperature and size independent dielectric function (dashed lines) is given.

The absorption behaviour at long wavelengths of a spherical iron grain with $a=0.01~\mu{\rm m}$ is dominated by the electric dipole absorption if the size dependence is taken into account. This is also true for the spheroids as the contribution of the electric dipole absorption only increases with axis ratio while that of the magnetic dipole absorption is assumed to be constant.
	 
As can be seen in the {figure}, by considering the size dependence of the dielectric function, the absorption at long wavelengths is clearly enhanced in comparison to the calculation where this dependence is ignored. On the other hand a resonance absorption in the IR due to surface modes (best seen for a prolate grain with $a/b=100$) is reduced. 

The absorption properties at long wavelengths is qualitatively different if the
magnetic dipole absorption becomes important, which is the case of larger iron grains. The behaviour should be similar to the case shown in {Fig.}~\ref{fig_abs_spheroids} where the size and temperature dependence of the dielectric function is ignored. As can be seen in the figure, at long wavelengths the absorption behaviour of the sphere is {then} dominated by the magnetic dipole absorption. 
Because of the assumption of magnetic dipole absorption,
the absorption efficiency at those wavelengths {is almost insensitive} to the axis ratio $a/b$ unless the electric dipole absorption becomes important. 

The absorption behaviour in the optical on the other hand increases with increasing axis ratio as long as the ratio is small. Therefore, slightly prolate or oblate {small} iron grains with a volume corresponding to a sphere with $r_v>0.015~\mu{\rm m}$ can absorb optical light more efficiently than spheres but show a similar cooling behaviour. If the grains are heated by the ISRF for example this results in an increase of the grain temperature with increasing axis ratio (Sect.\ref{equiltempspheroids}).




  \section{Iron grains in the ISM}

  \label{iron_ISM}
As an example we considered spherical and small spheroidal iron grains which are heated by the ISRF. For the radiation field we have taken the energy density in the solar neighbourbood as derived by Mathis, Mezger \& Panagia (MMP \cite{MMP82}), scaled by a factor $\chi$:
\begin{equation}
  u_{\lambda}= \chi \left(u_{\lambda}^{UV_\odot}+\sum_{i=2}^4 W_i \frac{4\pi}{c}B_{\lambda}(T_i)\right)
  +\frac{4\pi}{c}B_{\lambda}(2.7~{\rm K})
\end{equation}
 with dilution factors $W_2=10^{-14}$, $W_2=10^{-13}$, $W_3=4\times 10^{-13}$ and the 
 corresponding temperatures $T_2=7500$~K, $T_3=4000$~K and $T_4=3000$~K. $u_{\lambda}^{UV_\odot}$ is 
 the energy density of the ultraviolet photons in the solar neighbourhood (MMP \cite{MMP82}, Tab. C1).
 The maximum energy of the photons is 13.6~eV.

 \subsection{Grain temperature}

First we ignore the temperature variation in the case of very small grains and characterise all iron grains with single temperatures $T_{\rm eq}$ where the heating is equal to the cooling of the grain:
  \begin{equation}
    \int \mathrm{d}{\lambda}\,c\,u_{\lambda}\left<C_\mathrm{abs}\right> =
    \int{\rm d}\lambda\, 4\,\pi B_{\lambda}(T_\mathrm{eq})\, \left<C_\mathrm{em}\right>,
  \end{equation}
where $c$ and $B_{\lambda}(T_\mathrm{eq})$ are the velocity of light and the Planck function. 
$\left<C_\mathrm{abs}\right>$ and $\left<C_\mathrm{em}\right>$ are the absorption cross sections averaged over the orientations of the grain with $\left<C_{\rm em}\right>=\left<C_{\rm abs}\right>$. In Sect.~\ref{sect_flickering} we consider the temperature variation of small iron grains due to stochastic heating.

  \subsubsection{Equilibrium temperatures of iron spheres}
  \label{spheres_equiltemp}

\begin{figure}[htbp]
    \includegraphics[width=\hsize]{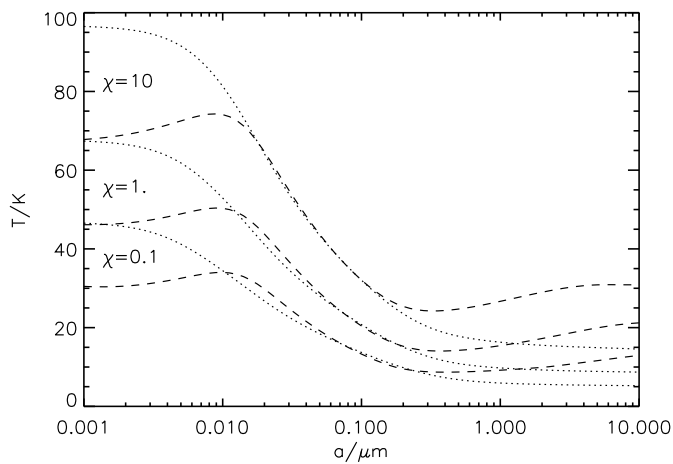}
    \caption{\label{equiltemp}
      Equilibrium temperatures of spherical iron grains in the ISM
      for three different radiation strengths ($\chi=0.1$, $\chi=1$ and $\chi=10$).
      Dashed and dotted lines correspond to calculations in which we included and ignored
      the temperature and size dependence of the dielectric function.}
  \end{figure}

  For the 
  photon density $u_{\lambda}$ we assumed three different strengths $\chi=0.1$ $\chi=1$ and $\chi=10$.
  For comparison we also derived the equilibrium temperatures for spheres where we 
  assumed a dielectric function which is independent of temperature and size. 
  The derived equilibrium temperatures of iron grains with radii ranging from 10~\AA{}
  to 10~$\mu{\rm m}$ are shown in Fig.~\ref{equiltemp}. 

  As mentioned by Chlewicki \& Laureijs (\cite{Chlewicki88}),
  the equilibrium temperature of iron grains in the ISM is a strong function of grain size.
  Where the magnetic dipole effect becomes important at $a\approx 0.015~\mu{\rm m}$ the
  temperature falls from relatively high to low values which are close to temperatures
  of silicate or graphite grains. 
  The strong temperature variation with grain size is also expected for collisionally heated
  iron spheres (see Fischera et al. \cite{fischera02}).

  The grain temperatures are so low that the collision rate of the free electrons is 
  dominated by surface scattering
		(see Fig.~\ref{crit_size}). 
  As a consequence of the absorption behaviour in the IR
  (Sect. \ref{abs_spheres}) the cooling of the smallest grains increases while the cooling
  of larger grains becomes less efficient. Therefore, if the size dependence is taken into account,
  the temperature of smallest iron spheres is lower and the temperature of
  larger iron spheres higher compared with the case where the size dependence of the dielectric
  function has been ignored. We mention that the cooling of the smallest grains could 
  be even stronger
  if the dielectric function in the IR were lower than assumed here. 


\subsubsection{Equilibrium temperatures of spheroids}

  \label{equiltempspheroids}
\begin{figure}[htbp]
    \includegraphics[width=\hsize]{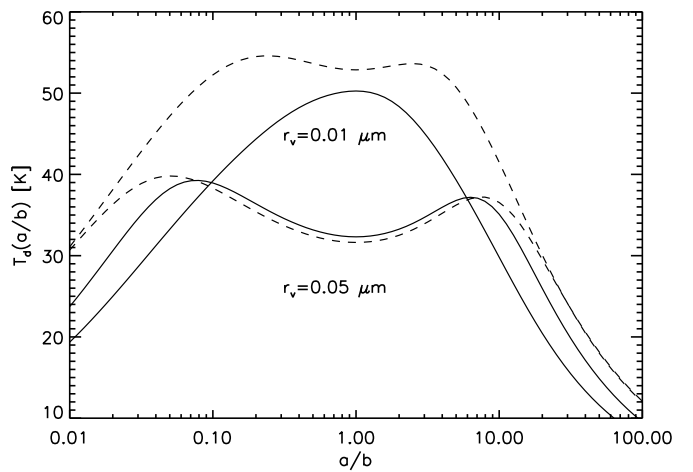}
    \caption{
      \label{temp_spheroids}
        Approximate equilibrium temperatures of spheroidal iron grains
        with a volume equal to a volume of a sphere with radius
    $r_{\rm v}=0.01~\mu{\rm m}$ and $r_{\rm v}=0.05~\mu{\rm m}$ as
    function of the ratio $a/b$. The
    solid and the dashed line correspond to calculations where the
    dependence of the dielectric function on grain size and temperature
        has been taken into account and has been ignored, respectively.
        }
  \end{figure}

For comparison we considered spheroids with a radius $r_{\rm v}=0.01~\mu{\rm m}$ and $r_{\rm v}=0.05~\mu{\rm m}$. The derived temperatures of these iron grains are shown in Fig.~\ref{temp_spheroids}. We also give the equilibrium temperatures of iron spheroids for a dielectric function ignoring the temperature or size dependence. The temperatures for extremely elongated or flattened iron grains, in particular for the grain with $r_v=0.05~\mu{\rm m}$, should be taken as approximate values because of the fact that in this case the electric dipole approximation in the UV gives only a poor representation of the actual absorption behaviour.

Taking the size and temperature dependence of the dielectric function into account, the two sizes considered show qualitatively a very different temperature behaviour. We found that the temperature of 
a spheroid with a radius $r_{\rm v}=0.01~\mu{\rm m}$ becomes smaller with an increasing axis ratio. This behaviour is therefore the same found for spheroids of other grain compositions
(Voshchinnikov et al, \cite{Vosh99}). On the other hand the temperature of an iron grain with $r_{\rm v}=0.05$ increases first before the temperature decreases for larger axis ratios. This is expected from the discussion at the end of Sect.~\ref{abs_spheroids}. 

  At extreme values of the ratio $a/b$ the equilibrium temperatures of
  spheroids with $r_{\rm v}=0.05~\mu{\rm m}$ are higher than for
  spheroids with $r_{\rm v}=0.01~\mu{\rm m}$. This can be
  explained by the higher collision rate due to surface scattering 
  in the smaller grain leading to a higher electric dipole absorption.
  
  By including the size and temperature dependence of the dielectric
  function the temperatures of spheroids with $r_{\rm v}=0.01~\mu{\rm
  m}$ are up to $\sim 10$~K lower than the temperatures where this dependence is
  ignored. If the dependences are not taken into account
  the temperature behaviour would qualitatively be
  similar to the larger grains showing an increase of the temperature
  for intermediate values of the ratio $a/b$ first and a decreasing 
  temperature for more oblate and prolate shapes.

  \subsubsection{Flickering small iron grains}
\label{sect_flickering}
  We derived the probability distribution of the temperatures of small grains
  using a numerical method described by Guhathakurta \& Draine (\cite{Guhathakurta89}).
  To calculate the thermal energy of the iron spheres 
  we made use of tabulated values of the heat capacity 
  {up to a temperature of \mbox{298 K} (`American Institute of Physics Handbook', \cite{AIP72}). At higher 
  temperatures an analytical expression of the heat capacity is used (Chase \cite{Chase98}). For this,}
  we assumed iron to be in the $\alpha$-$\delta$-phase.

\begin{figure}[htbp]
    \includegraphics[width=\hsize]{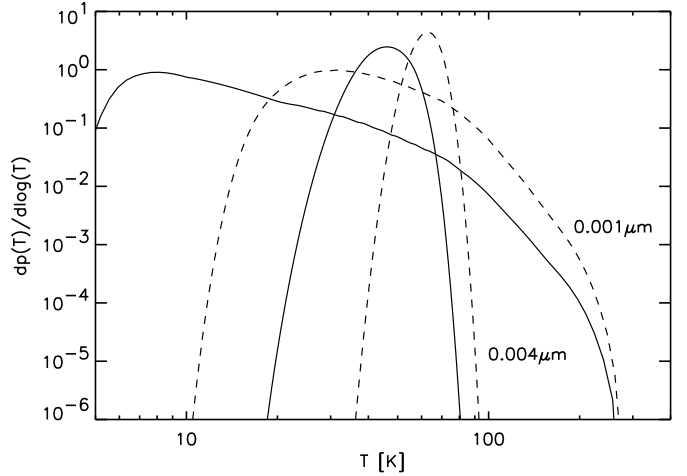}
    \caption{
      \label{flickering}
      Temperature distribution of spherical iron grains with radius 0.001 and $0.004~\mu{\rm m}$
      heated by the ISRF (MMP). Solid and dashed lines correspond to calculations in which we included
      and ignored the dependence of the dielectric function on temperature and grain size.}
  \end{figure}

  The temperature distributions of iron grains with radius $a=10~$\AA{} and $a=40~$\AA{} heated by the ISRF ($\chi=1$)
  are shown in \mbox{Fig. \ref{flickering}} as ${\rm d}p(T)/{\rm d}\log(T)$. 
  For comparison we also included in this figure 
  the theoretical temperature variation
  of small spheres with a dielectric function of bulk iron at room temperature. If the size dependence
  of the collision rate is taken into account the probabilities of high temperatures are partly
  strongly reduced while the probabilities of low temperatures is clearly increased.
  Iron spheres with $a=10~$\AA{} for example should have
  temperatures lower than \mbox{10 K} most of the time. 
  In contrast a model where the size dependence of the dielectric function is ignored 
  predicts for the same 
  grain size temperatures higher than \mbox{10 K} with a most probable temperature of \mbox{30 K}.
  
  The maximum temperature very small grains reach is roughly the temperature
  where the thermal energy is equal to the maximum photon energy of
  13.6~eV. Because of the high heat capacity of iron grains 
  the maximum temperature is clearly lower 
  in comparison with other grain species like silicate or graphite grains.
  By using the heat capacity given by Dwek (\cite{Dwek86}) graphite
  grains with a radius of $10$~\AA{} attain a maximum temperature of above
  400~K (Fischera \cite{fischera00}). In contrast the maximum temperature of iron grains with the
  same size is less than 300~K.

\subsection{The SED of spherical iron grains in the ISM}

From the analysis of the extinction curve Mathis, Rumpl \& Nordsieck (\cite{MRN77}) proposed
a power law ${\rm d}n(a)\propto a^{-k}\,{\rm d}a$ with $k\approx 3.5$ for the
grain size distribution in the ISM. The minimum and maximum grain size as the power of the 
size distribution depends somewhat on the assumed grain composition.
For the iron grains we considered a maximum grain size of $a_{\rm max}=0.1~\mu{\rm m}$. To study the effect of stochastic heated grains we assumed as minimum grain size different radii 10, 20 and 40~\AA{}.

\begin{figure}
  \includegraphics[width=\hsize]{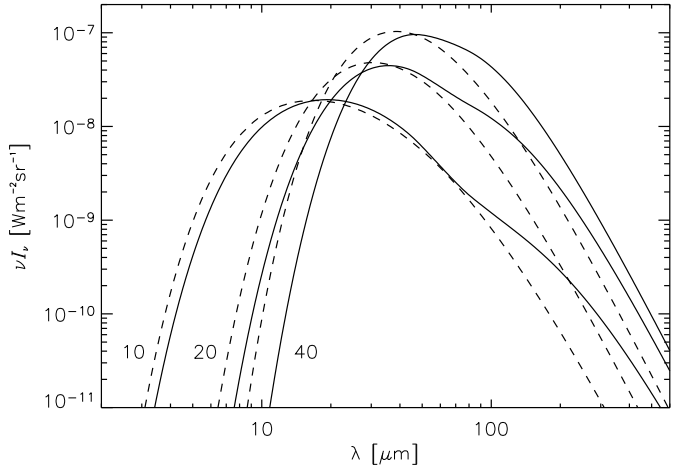}
  \caption{\label{spectra_single}
  Emission spectra of individual small iron grains. The sizes are chosen to be 10, 20, and 40~\AA{}. 
  The spectra shown as solid lines correspond to calculations in which the size and temperature
  dependence of the dielectric function has been taken into account. The spectra shown as dashed lines 
  correspond to calculations in which this dependence has been ignored.
  }
\end{figure}

{The mean intensity emitted at the surface of a flickering grain with radius $a$ is given by
\begin{equation}
  I_\nu(a) = \int{\rm d}T\,p(a,T)\,B_{\nu}(T)\,Q_{\rm em}(a,\nu,T),
\end{equation}
where $B_{\nu}(T)$ is the Planck-function and $p(a,T)$ the probability 
that the grain with radius $a$ has the temperature $T$. The spectral energy distribution 
(SED) $\nu I_\nu$ of three flickering iron grains of different sizes 
is shown in Fig.~\ref{spectra_single}. By including the size and temperature 
dependence of the dielectric function
the emission spectrum is shifted to longer wavelengths. This is expected as these grains cool 
more efficiently and therefore, as shown in the previous section, have higher probabilities at 
low but lower probabilities at high temperatures.}

{The total flux density of a distribution of spherical iron grains at distance $D$ is given by:
\begin{equation}
  F_{\nu}=\frac{1}{D^2}\int {\rm d}a \,A\,a^{-3.5} \,a^2 \pi I_{\nu}(a),
\end{equation}}
where we normalized $A$ to the total dust mass $M_{\rm d}$ of iron grains 
with the density $\rho_{\rm d}=7.87~{\rm g/cm^3}$:
\begin{equation}
 A=M_{\rm d}/\int{\rm d}a\,a^{-3.5}\frac{4}{3}\pi a^3 \rho_{\rm d}.
\end{equation}

\begin{figure}[htbp]
    \includegraphics[width=\hsize]{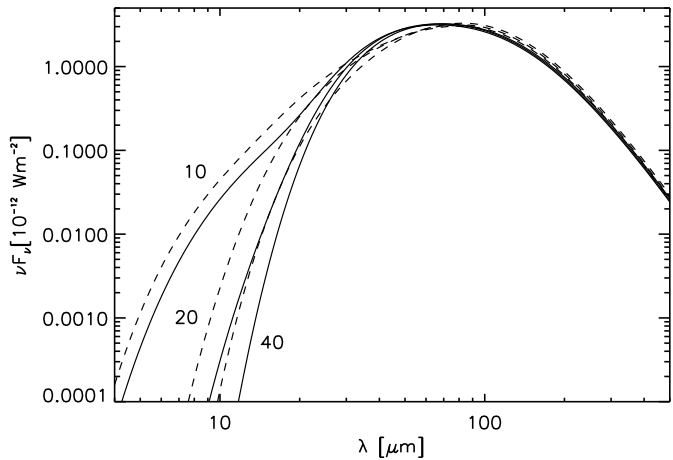}
    \caption{\label{SED}Spectral energy distribution (SEDs) of spherical iron grains with a mass 
	of $1~M_{\sun}$ at a distance of 1kpc heated by the ISRF (MMP). The grains are 
	assumed to have a power law $\mathrm{d}n(a)\propto a^{-3.5}\,{\rm d}a$ with a maximum 	grain size $a_\mathrm{max}=0.1~\mu{\rm m}$
      	and different minimum grain sizes $a_\mathrm{min}$
      	of $10$, $20$ and $40~\mathrm{\AA{}}$. SEDs in which calculation the
      	temperature and size dependence of the dielectric function was included are shown as 
	solid lines. Dashed lines correspond to SEDs in which this dependence has been ignored.}
  \end{figure}

The SED $\nu F_{\nu}$ of iron grains of $1~M_{\sun}$ at 1~kpc is shown in \mbox{Fig. \ref{SED}}.
For comparison we give the SEDs for the calculation including and ignoring the size and temperature dependence of
the dielectric function. Due to the large variation of the equilibrium temperatures the SED is relatively broad.
In comparison to graphite or silicate grains the SED is less effected by flickering grains. A clear increase at
the blue side of the spectrum requires grains with radii smaller than $a_{\rm min}\approx 20~$\AA{}.
The inclusion of the size and temperature dependence of the dielectric function 
reduces the emission at the blue side which is contributed by
stochastically heated small grains. The peak of the emission on the other hand is shifted towards
shorter wavelengths. At long wavelengths we obtained for the two different cases of the
dielectric function almost the same emission spectrum.

\section{Discussion}

The study of the temperature behaviour of metallic grains in the ISM requires 
a realistic model of the ir optical properties. The grains emit mainly in the IR,
in which range the absorption is strongly affected by the free electrons. This
leads to an absorption that depends on temperature and, in cases where
the free path of the electrons is larger than the grain size, especially on the grain size.
A simplification can easily cause bigger uncertainties
in the equilibrium temperatures of spherical and spheroidal grains
or in the probability distribution of temperatures from flickering small grains.
      
To derive the cooling behaviour of iron grains Tabak and Straitiff (\cite{Tabak83}) and Tabak (\cite{Tabak87}) only considered the temperature but no size dependence of the optical properties. In addition it was assumed that the grain absorption is sufficiently described by the electric dipole approximation. Both simplifications lead to absorption coefficients which are too low. Therefore, equilibrium temperatures in the range of 100~K for iron grains heated by the ISRF as presented by Tabak \& Straitiff (\cite{Tabak83}) and also Tabak (\cite{Tabak87}) should be in general too high.

The temperatures of iron spheroids with $r_{\rm v}=0.01~\mu{\rm m}$ as given by Voshchinnicov et al. (\cite{Vosh99}) show qualitatively the peculiar behaviour we derived only for larger spheriods where the absorption behaviour at long wavelengths of the corresponding sphere is dominated by the magnetic dipole absorption. The peculiar curve of spheroidal grains with $0.01~\mu{\rm m}$ should disappear if the size dependence of the dielectric function is taken into account.

\subsection{Uncertainties in grain temperatures}    

\subsubsection{Uncertainties due to impurities and defects}
\label{sectdiscuss1}
It is likely that the iron particles in astrophysical environments show a high degree
of impurities and lattice defects. And it is possible due to the surface tension that the lattice
structure becomes more disordered towards smaller grains. Both impurities and defects lead to a higher scattering rate inside the solid in comparison to bulk iron. As shown this effect should be negligible for iron grains with radii smaller than $0.25~\mu{\rm m}$ if $\zeta\le100$. 
Larger grains on the other hand will be affected. 

We found by considering iron spheres heated by the ISRF that the grain temperature as a function of radius shows a minimum at $\sim 0.3~\mu{\rm m}$ and that the temperature increases towards even larger iron grains. This is caused by the fact, as is shown in Sect.~\ref{approx} and Sect.~\ref{idealised_dipole}, that the absorption behaviour of big iron grains is dominated by the magnetic dipole absorption (Eq.~\ref{Qmagndipol3}) which decreases towards larger grains if the scattering rate is dominated by scattering events at the grain surface and therefore given by $\Gamma=v_{\rm F}/a$. In the case of less pure and disordered iron grains the collision rate will be dominated by scattering events at impurities and defects so that the absorption coefficient becomes essentially constant with grain size. It is therefore possible that the increase in grain temperature will disappear if iron grains in astrophysical environments are considered.
        
\subsubsection{Uncertainties in the dielectric function}
  A better description of the iron temperatures requires a better understanding
  of the optical properties in the IR. To include the temperature and size dependence
  of the dielectric function
  we assumed that the free electron part is given by a single Drude term. As a
  consequence we obtained high values in the IR which were attributed to the optical response
  of the bound electrons and
  were assumed to be independent of temperature and size.
  But measurements by Weaver et al. (\cite{Weaver79}) of the optical properties 
  of bulk iron at temperatures in the region of 140~K 
  indicate that the contribution of bound electrons to the optical behaviour
  is strongly decreasing at energies lower than 0.8~eV. 
        
   In this case, as can be seen in Fig.~\ref{epsilon_ir}, the IR-values of the dielectric function of very small grains could be lower and their absorption efficiency therefore higher than assumed here. If the part of the bounded electrons would be negligible so that the dielectric function is determined by the free electron part, the absorption or cooling efficiency of an iron grain with $a=0.001~\mu{\rm m}$ could increase by almost one order of magnitude.
It is therefore possible that the probabilities at high temperatures of flickering small iron grains
is even smaller than the probabilities presented here.

\section{Summary}

  The optical properties of small spherical and spheroidal
  iron grains were derived using a self consistent model of the dielectric function
  including its dependence on size and temperature. We have shown that
  the limitation of the free path of the electrons due to the size of the grain
  has a non-negligible effect on the absorption behaviour of small spheres and spheroids.
  We applied the estimated absorption coefficients for small spherical and spheroidal
  iron grains in the ISM which are heated by the ISRF and derived the emission
  spectrum of stochastically heated small iron spheres assuming as grain size
  distribution a power law ${\rm d}n\propto a^{-3.5}\,{\rm d}a$ as proposed for
  grains in the ISM and chosing the radius of the largest grain as $0.1~\mu{\rm m}$.
  The main results for the temperature behaviour of small iron grains are:

  \begin{itemize}
  \item 
   The equilibrium temperatures of pure iron spheres heated by the ISRF
			are strongly size dependent. 
   The highest equilibrium temperature of $\sim 50$~K was
		 found for iron spheres with radius $\sim 0.01~\mu{\rm m}$ and the lowest temperature
   of $\sim 14$~K for iron spheres with $a\approx 0.3~\mu{\rm m}$. 
{The temperature behaviour of larger iron grains is likely to be effected by impurities and defects.    If these grains are relatively pure the temperature should increase with size. Otherwise their temperature should be size independent.}
  \item
   In case of iron spheroids with a volume corresponding to a sphere with radius
	  $r_{\rm v}=0.01~\mu{\rm m}$ the temperature decreases 
   for grains with oblate and prolate shape. 
  \end{itemize}

  By considering the size dependence of the dielectric function we found in comparison
  to calculations where this dependence is ignored:
  \begin{itemize}
   \item The temperatures of small spheroidal grains ($r_{\rm v}\approx 0.01~\mu{\rm m}$) 
   should be significantly lower.
    \item Absorption in the case of small grains 
          (where electric dipole absorption is important) increases while
          absorption in the case of bigger grains 
          (where magnetic dipole absorption is important) decreases.
  \item 
   The absorption at resonance
   frequencies caused by shape effects of strongly elongated or flattened
   iron grains with $r_{\rm v}=0.01~\mu{\rm m}$ is significantly reduced. 
   The absorption at 
   IR-wavelengths beyond the resonance wavelengths increases significantly.
  \item 
   Due to better cooling the probabilities of high temperatures of small
   flickering grains should be lower and the probabilities of low temperatures
   significantly higher.
  \item 
   The contribution of flickering grains to the SED at short wavelengths is 
   reduced.
  \end{itemize}

\begin{acknowledgements}
I am thankful for the support provided by the Max-Planck-Institut f''ur Kernphysik in Heidelberg and the Research School of Astronomy \& Astrophysics of the Australian National University in Canberra. I acknowledge the financial support through the ARC Discovery project DP0208445.
  I would like to thank Prof. Kr\"atschmer for fruitful discussions and Prof. Dopita, 
  Roberto de Propris, and Helmut Jerjen for useful comments that improved the manuscript. 
\end{acknowledgements}

\appendix

\section{Analytical approximation at long wavelengths}
{ 
\label{approx}
Here we will discuss in some detail the absorption behaviour of metallic grains using the electric and magnetic dipole approximation. To derive approximations it is assumed that the dielectric function is dominated by the free electron gas and therefore by $\epsilon^f$.
}
 The cross section at wavelengths with $2\pi a/\lambda\ll 1$ is given by (see e.g. Draine \& Lee \cite{DraineLee84}):
  \begin{eqnarray}
    \label{dipole_approx}
    C_{\rm abs} = C^e_{\rm abs}+C^m_{\rm abs}=\frac{4\pi\omega}{c}{\rm Im}\left(\alpha^e+\alpha^m\right),
  \end{eqnarray}
  where ${\rm Im}\left(\alpha^e+\alpha^m\right)$ is the imaginary part of 
  the electric and the magnetic {polarizability}. In the case of spheres these are:
  \begin{eqnarray}
    \label{el_dipole}
    \alpha^e &=& a^3 \frac{\epsilon-1}{\epsilon+2},\\
    \label{magn_dipole}
    \alpha^m &=& -\frac{a^3}{2}\left(1+\frac{3}{y}\cot{y}-\frac{3}{y^2}\right)~\mathrm{with} ~
       y^2=\epsilon\left(\frac{\omega a}{c}\right)^2.
  \end{eqnarray}

If the contribution of the magnetic dipole absorption is negligible the absorption coefficient is known to be given by:
  \begin{equation}
    \label{Qeldipol}
    Q_\mathrm{abs}\approx Q_\mathrm{abs}^e=\frac{\omega^2 12
  a\Gamma}{c\omega_p^2}.
  \end{equation}
Therefore, the absorption is proportional to $\lambda^{-2}$ and independent of grain size where
the collision rate is given by $\Gamma=v_F/a$. This is the case for very small iron grains at wavelengths larger $\sim 60~\mu{\rm m}$. At lower wavelengths the absorption is dominated by $\epsilon^b$ which causes a reduced absorption with $Q_{\rm abs}\propto a/\lambda^2$. 

With increasing grain size the magnetic dipole absorption begins to dominate the absorption behaviour at long wavelengths. In the region $|y| \ll 1$ and $\omega \ll \Gamma$, so that $\epsilon^f\approx i\omega_p^2/\omega\Gamma$, the absorption coefficient due to induced eddy currents in the metal spheres can be simplified to the expression:
  \begin{equation}
    \label{Qmagndipol1}
    Q_\mathrm{abs}^m\approx \frac{12}{90}\left(\frac{\omega
  a}{c}\right)^3\frac{\omega_p^2}{\omega \Gamma}.
  \end{equation}
Taking $\Gamma=v_F/a$ the absorption is proportional to $a^4/\lambda^2$. Comparing the
approximations of the electric (Eq. \ref{Qeldipol}) and the magnetic dipole absorption (Eq. \ref{Qmagndipol1}) both terms are equal at a grain size of $a\approx 0.015~\mu{\rm m}$. This results in the strong increase of the absorption at long wavelengths seen in Fig. \ref{optprop} with grain size from 0.01 to $0.1~\mu{\rm m}$.

In the region $\omega \ll \Gamma$, where $|y| \gg 1$, the absorption can be described by:
  \begin{equation}
    \label{Qmagndipol2}
    Q_\mathrm{abs}\approx Q_\mathrm{abs}^m\approx
    \frac{3\sqrt{2}}{\sqrt{\epsilon_2}}-\frac{3}{\epsilon_2}=
    \frac{3\sqrt{2\omega\Gamma}}{\omega_p}
    \left(1-\frac{\sqrt{\omega\Gamma}}{\sqrt{2}\omega_p}\right).
  \end{equation}
The absorption is therefore proportional to $1/\sqrt{\lambda}$ for a fixed collision rate $\Gamma$ and proportional to $1/\sqrt{a\lambda}$ if $\Gamma=v_F/a$. For a pure cold iron sphere with $a=1~\mu{\rm m}$ this behaviour occurs only at very long wavelength with $\lambda >1000~\mu{\rm m}$  (Fig.~\ref{absdipol}).

The plateau with constant absorption coefficients at long wavelengths is caused by the magnetic dipole absorption in the region $|y|\gg 1$ and 
  $\omega_p\gg \omega \gg \Gamma$ where $\epsilon^f\approx 
  -\omega^2_p/\omega^2(1-i\Gamma/\omega)$. 
In this range the absorption coefficients are approximately given by:
  \begin{equation}
    \label{Qmagndipol3}
    Q_\mathrm{abs}^m\approx 
      \frac{3\Gamma}{\omega_p}\left(1-\frac{2c}{a\omega_p}\right).
  \end{equation}  
Bigger iron grains in this parameter range absorb less efficiently than smaller grains do if $\Gamma=v_F/a$. {As discussed in Sect.~\ref{sectdiscuss1} this behaviour implies relatively pure and ordered iron grains with $\zeta\approx1$. In case of less pure or disordered iron spheres the absorption efficiency may be independent of grain size.}

For completeness we also give the absorption behaviour for the region
$\omega_p\gg \omega \gg \Gamma$ and $|y|\ll 1$:
  \begin{equation}
    \label{Qmagndipol4}
    Q_{\rm abs}^m \approx \frac{12}{90}\frac{\omega_p^2 a^3 \Gamma}{c^3}.
  \end{equation}
The magnetic dipole absorption in this region is again independent of wavelength {and is proportional to $a^2$ if the collision rate is given by $\Gamma=v_F/a$. But  this behaviour is not very prominent in case of iron spheres. Fig.~\ref{optprop} shows only a flattening of the absorption coefficients of a grain with a radius of $0.1~\mu{\rm m}$ between 10 and 100~$\mu {\rm m}$.}

\section{Idealised dipole absorption of iron}
\label{idealised_dipole}

 \begin{figure*}[htbp]
    \includegraphics[width=0.5\hsize]{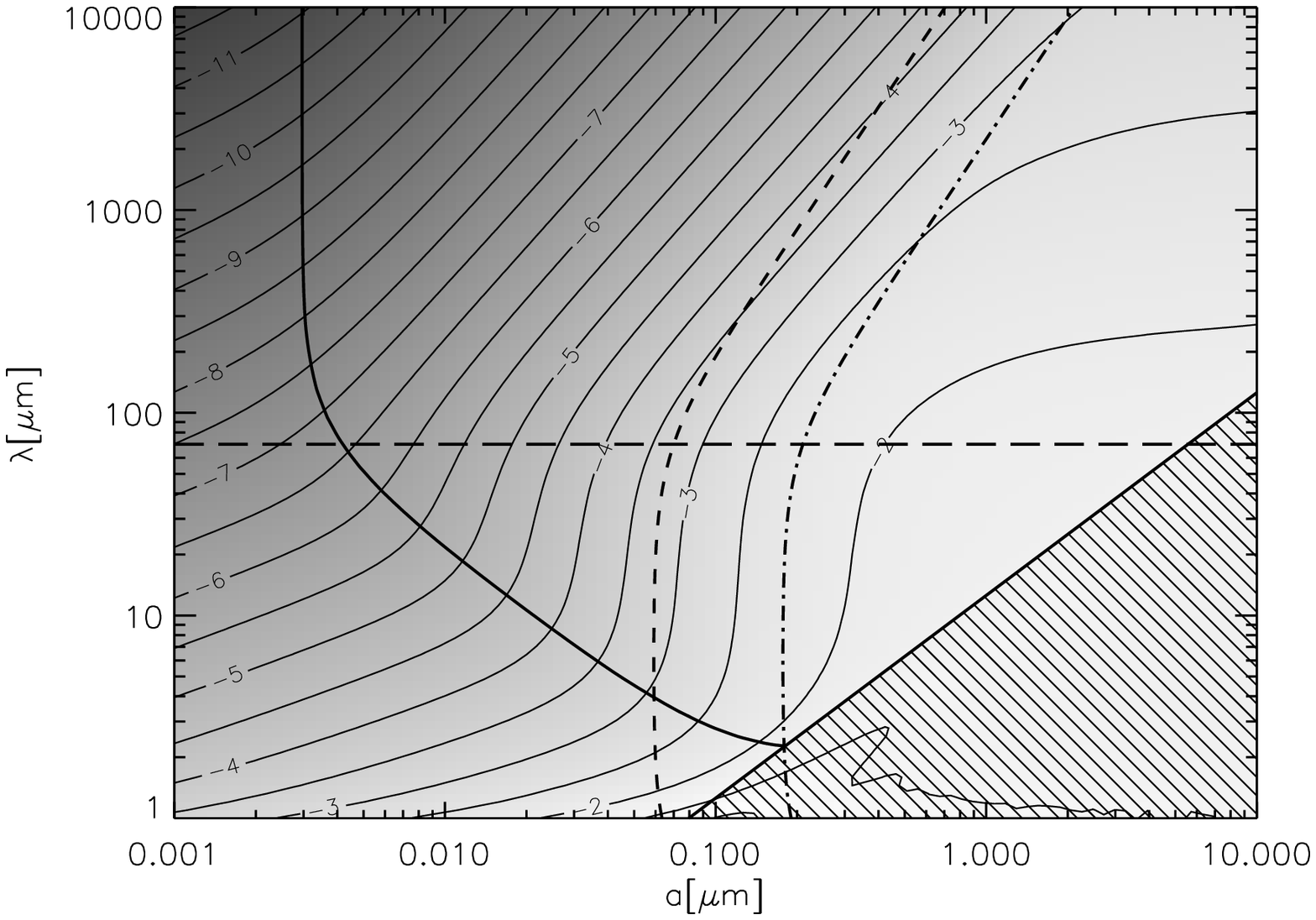}
    \includegraphics[width=0.5\hsize]{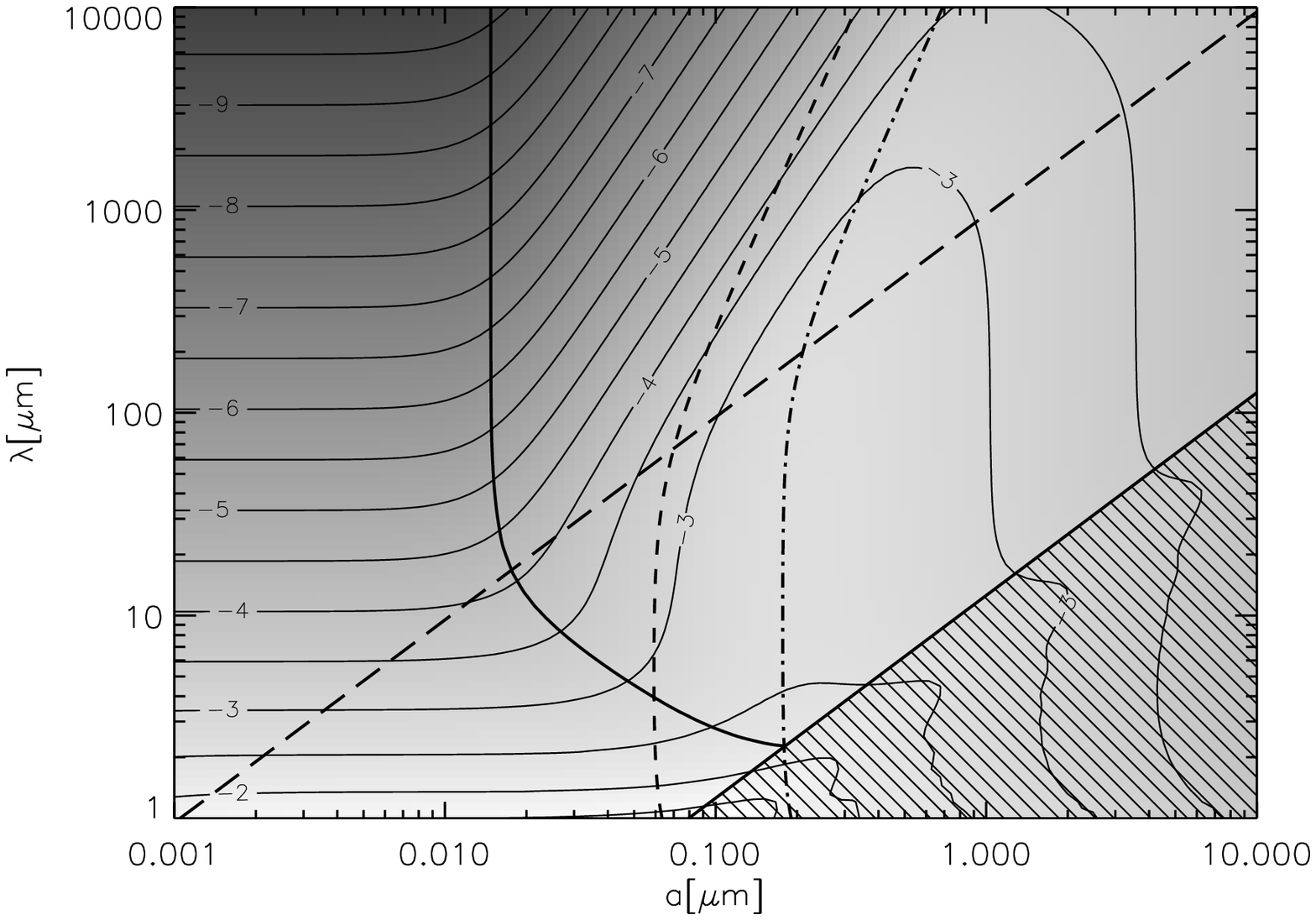}
    \caption{\label{absdipol}
    Absorption coefficients $Q_{\rm abs}(a,\lambda)$ of metallic iron spheres
    derived approximating the dielectric function $\epsilon$ through the 
    free electron part $\epsilon^{\rm f}$. In the left hand figure the collision rate of 
    the free electrons is taken to be the same as for bulk iron at 298~K. In the right 
    hand figure it is assumed that the collision rate is dominated by  
    scattering events at the grain surface.  
    The absorption coefficients
    are shown as contours in logarithmic scale (thin solid lines)
    where the difference between the lines is taken to be
    $\Delta \log_{10}(Q_{\rm abs}(a,\lambda))=0.5$. The labels correspond to 
    $\log_{10}(Q_{\rm abs}(a,\lambda))$.
    The hatched region shows where the
    absorption behaviour deviates strongly from the dipole absorption 
    ($2\pi a/\lambda > 0.5$). The thick solid line visualises where
    the electric dipole absorption is equal to the magnetic dipole absorption.
    Where the frequency $\omega$ of the photons is equal to the
    collision rate $\Gamma$ of the free electrons in the grain is shown
    as long dashed line. The short dashed and dashed dotted line correspond
    to absorption coefficients where $|y|=1$ and $|y|=3$, respectively.}
\end{figure*}
 
To visualise the effect of the limitation of the mean free path lengths of the free electrons on the cooling behaviour of iron spheres it might be useful to approximate the dielectric function by a single Drude term using the parameters as given in Table~\ref{drudeparameter}. The derived absorption coefficients are shown in Fig.~\ref{absdipol} as function of wavelength and grain size. For comparison the absorption coefficients have been calculated for two different assumptions for the collision rate $\Gamma$. In the left hand figure no correction to the measured dielectric function (at room temperature) has been made. In the right hand figure it is assumed that the collision rate is dominated by surface scattering which may be appropriate to describe the situation of cold pure iron grains in the ISM which are heated by the ISRF. 

If no correction is applied, the magnetic dipole absorption would be the dominant absorption at long wavelengths for grains with radii larger than 3~nm. The limitation of the free path of the free electrons by the grain size shifts this value to grains with radii larger than $\sim 0.015~\mu{\rm m}$. The absorption increases almost for all wavelengths with grain size when no size dependence of the dielectric function has been considered (Fig.~\ref{absdipol}, left hand side). When this dependence taken into account, the absorption in the IR shows a maximum for grains with a radius in the range of $\sim 0.3~\mu{\rm m}$ (Fig.~\ref{absdipol}, right hand side), 
which is the reason for the lowest temperatures of these grains in the ISM as derived in Sect.~\ref{spheres_equiltemp}. The decrease of the absorption with grain size for grains larger than $0.3~\mu{\rm m}$ extends also into the region where the absorption has to be described using the Mie-formalism (hatched region). 

The approximations of the magnetic dipole absorption given in 
Sect.~\ref{abs_spheres} are valid in the regions left of $|y|\approx 1$ and 
right of $|y|\approx 3$ (shown in Fig.~\ref{absdipol} as short dashed and dashed 
dotted lines) considering $\omega \gg \Gamma$ and $\omega \ll \Gamma$.

{The absorption behaviour is shown for non-magnetic iron grains to demonstrate the different trends. As stated in the text, above $\lambda\approx 1000~\mu{\rm m}$ the absorption of iron grains might be dominated by magnetisation effects (Fig.~\ref{abstemp}).
At short wavelengths on the other hand the absorption behaviour of the iron grains additionally may be effected by the response of the bound electrons.}


\end{document}